\documentclass[10pt,a4paper,twocolumn]{article}
\usepackage[english]{babel}
\usepackage[left=0.7in,top=0.7in,bottom=0.9in,right=0.5in]{geometry}
\usepackage[svgnames,x11names]{xcolor}
\usepackage{caption}
\usepackage[caption=false]{subfig}
\usepackage{graphicx}
\usepackage{amsmath,amssymb}
\usepackage{ctable}
\usepackage{upgreek}
\usepackage{cite}
\usepackage[colorlinks=true
,urlcolor=blue
,anchorcolor=blue
,citecolor=blue
,filecolor=blue
,linkcolor=blue
,menucolor=blue
,linktocpage=true
,pdfproducer=medialab
]{hyperref}
\usepackage{breakurl}
\usepackage[noindentafter]{titlesec}
\titleformat{\section}
  {\normalfont\fontsize{12}{15}\bfseries}{\thesection}{1em}{}

\usepackage{listings}
\usepackage{hyphenat}
\newcommand\mcmule{{\sc McMule}}
\newcommand\moller{{M\o ller}}
\begin{document}
\twocolumn[{%
\begin{@twocolumnfalse}
\vspace{-3em}
\begin{flushright}
PSI-PR-21-15\\
ZU-TH 34/21\\
IPPP/21/14
\end{flushright}
\vspace{1em}

\begin{center}
{\Large\bf \moller{} scattering at NNLO}
\\[2em]
{
Pulak Banerjee$^{a}$,
Tim Engel$^{a,b}$,
Nicolas Schalch$^{c}$,
Adrian Signer$^{a,b}$,
Yannick Ulrich$^{d}$
}\\[0.1in]
{\sl ${}^a$ Paul Scherrer Institut,
CH-5232 Villigen PSI, Switzerland \\
${}^b$ Physik-Institut, Universit\"at Z\"urich, 
Winterthurerstrasse 190,
CH-8057 Z\"urich, Switzerland\\
${}^c$ Albert Einstein Center for Fundamental Physics, Institut f\"ur
Theoretische Physik,\\ Universit\"at Bern,
Sidlerstrasse 5, CH-3012 Bern, Switzerland\\
${}^d$ Institute for Particle Physics Phenomenology, University of
Durham, \\
South Road, Durham DH1 3LE, United Kingdom
}
\setcounter{footnote}{0}
\end{center}

\vspace{0em}

\begin{center}
\begin{minipage}{6in}
{\small We present a calculation of the full set of next-to-next-to-leading-order QED corrections to unpolarised \moller{} scattering. This encompasses photonic, leptonic, and non-perturbative hadronic corrections and includes electron mass effects as well as hard photon radiation. The corresponding matrix elements are implemented in the Monte Carlo framework \mcmule{} allowing for the computation of fully-differential observables. As a first application we show results tailored to the kinematics and detector design of the PRad~II experiment where a high-precision theory prediction for \moller{} scattering is required to achieve the targeted precision. We observe that the corrections become essential to reliably calculate the corresponding differential distributions especially in regions where the leading-order contribution is absent.}

\end{minipage}
\end{center}
\vspace{1em}
\end{@twocolumnfalse}}]

\section{\label{sec:intro}Introduction}
Recent developments at the low-energy precision frontier renewed interest in high-precision calculations in QED. In order to meet the experimental accuracy when using state-of-the-art lepton beams, next-to-leading-order (NLO) or even next-to-next-to-leading-order (NNLO) corrections in QED have become highly desirable. 

Electron-electron or \moller{} scattering is a prime example where a
high-precision theory prediction has become important.
As it is a ubiquitous process at electron beam lines, it
has been investigated in connection with luminosity
measurements~\cite{Schmidt:2017jby,Benisch:2001rr} and background
studies~\cite{DarkLight}, including a recent dedicated measurement at
very low energies to study electron mass
effects~\cite{Epstein:2019gjd}. 
Precise knowledge of QED
effects might also be useful for the MOLLER
experiment~\cite{MOLLER:2014iki} that searches for
parity violation as an indication towards new physics. The main motivation, however, is given by PRad~\cite{Xiong:2019umf,Gasparian:2020hog} to which we turn below.

All of these experiments rely on Standard Model theory predictions for \moller{} scattering. Since at low energies the corresponding radiative corrections are QED-dominated, a high-precision theory calculation is feasible. To account for non-trivial detector geometries and acceptances these experiments rely on Monte Carlo event generators that combine matrix elements to physical observables. At NLO accuracy in QED such Monte Carlo codes have been developed~\cite{Afanasev:2006xs,Akushevich:2015toa,Epstein:2016lpm}. While for most experiments this level of precision is sufficient, the situation is different for the planned PRad~II experiment~\cite{Gasparian:2020hog}, the upgraded version of its predecessor PRad~\cite{Xiong:2019umf}.

Both experiments measure the elastic scattering of electrons and protons to extract the proton charge radius. These efforts were triggered after a recent high-precision spectroscopy experiment using muonic hydrogen at the Paul Scherrer Institute~\cite{Pohl:2010zza,Antognini:1900ns} had measured a significantly smaller value for the proton charge radius compared to earlier results~\cite{RevModPhys.80.633}. The PRad and PRad~II experiments therefore play an important role in the resolution of this `proton radius puzzle'.

A key feature of the PRad experiments is the suppression of systematic uncertainties by normalising to a simultaneous measurement of \moller{} scattering. To be precise, the differential cross section for electron-proton scattering $(d\sigma/d\theta)^\mathrm{exp}_{ep}$ is extracted from the measured events $N_\mathrm{exp}$ and the theoretical prediction for the \moller{} cross section $(d\sigma/d\theta)^\mathrm{th}_{ee}$ via \cite{pradproposal}
\begin{equation}
    \left(\frac{d \sigma}{d \theta}\right)^\mathrm{exp}_{ep} 
    = \left[
    \frac{N_{\mathrm{exp}}(e^- p \rightarrow e^- p)} {N_{\mathrm{exp}}(e^- e^- \rightarrow e^- e^-)} \right] \left(\frac{d \sigma}{d \theta}\right)^\mathrm{th}_{ee}.
\end{equation}
It is therefore essential that the theory prediction for \moller{} scattering matches the experimental precision.

As alluded to before, the existing NLO Monte Carlo event generators were sufficient for PRad. This, however, is not the case for PRad~II that aims at improving the experimental precision significantly. It is therefore expected that missing higher-order QED corrections become the dominant source of systematic uncertainty~\cite{Gasparian:2020hog}.

We have therefore calculated the full set of NNLO QED corrections to \moller{} scattering. This includes not only photonic and leptonic corrections but also non-perturbative hadronic contributions. The corresponding matrix elements were implemented in the \mcmule{} framework, a Monte Carlo for MUons and other LEptons~\cite{Banerjee:2020rww}. This allows for the computation of arbitrary fully-differential observables.

This paper is organised as follows: we begin by briefly summarising
our calculational methods in Section~\ref{sec:calculation}. Next, we present results tailored to the kinematics and detector design of the PRad II experiment in Section~\ref{sec:results} and conclude in Section~\ref{sec:conclusion}.

\section{\label{sec:calculation}Calculation}

\begin{figure}[t]
\centering
\subfloat[Examples of photonic corrections at NNLO]{
         \includegraphics[width=.465\textwidth]{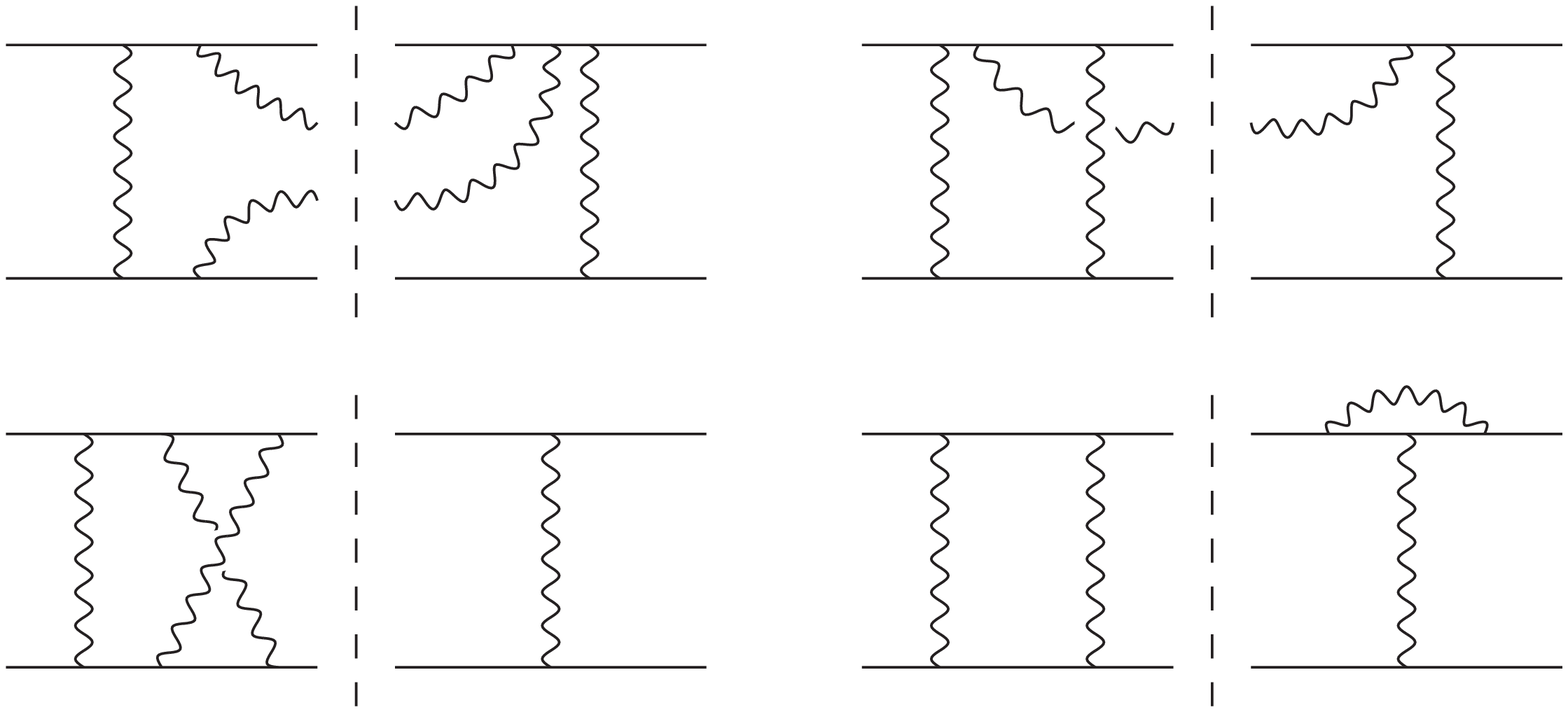}
         \label{fig:photon}
}\\
\subfloat[Examples of fermionic corrections at NNLO]{
         \includegraphics[width=.465\textwidth]{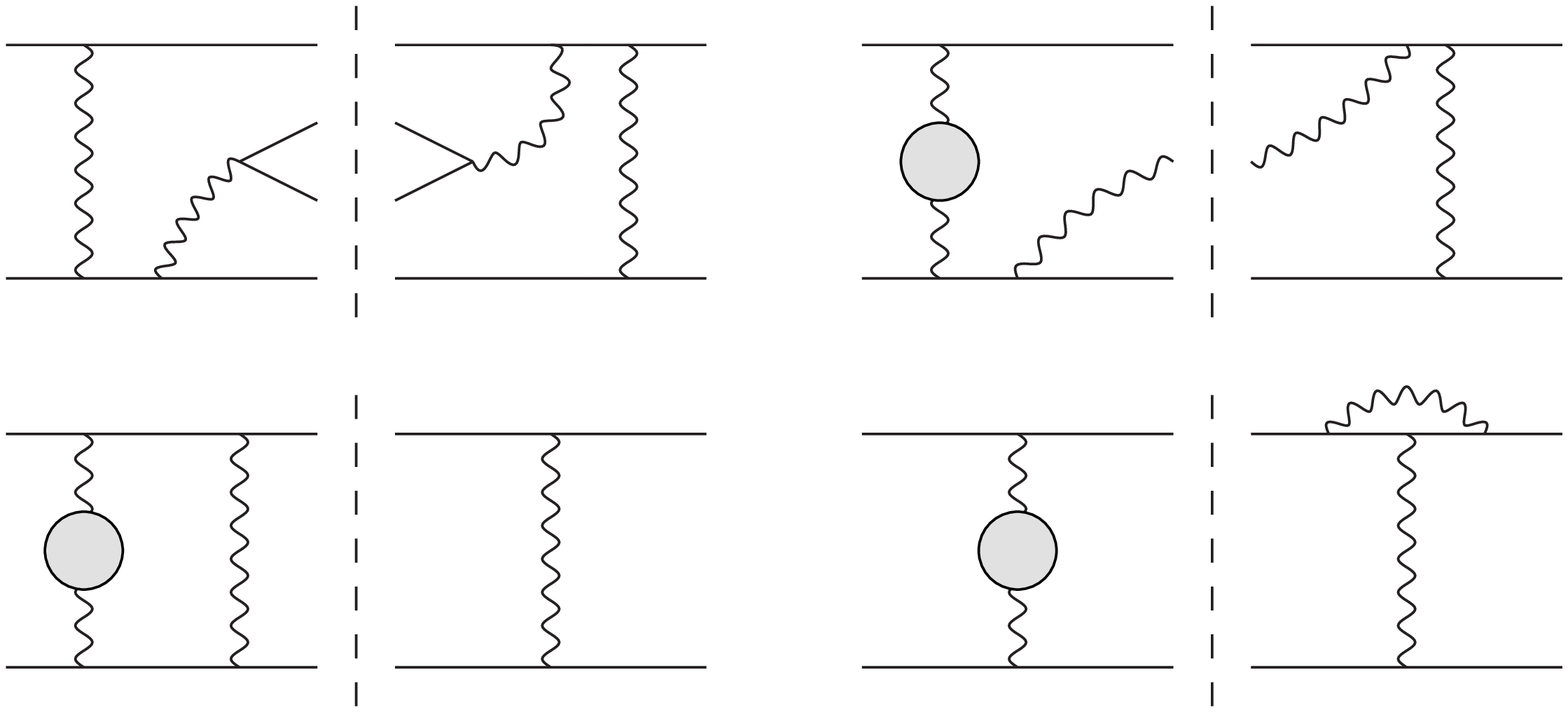}
         \label{fig:fermion}
}
\captionof{figure}{\label{fig:diags}
Representative contributions to the squared amplitude, double-real (top left), real-virtual (top right), and double-virtual involving the two-loop amplitude (bottom left) and one-loop amplitude squared (bottom right).}
\end{figure}

We have calculated the full set of NNLO QED corrections to the process
\begin{equation}
    e^-(p_1) e^-(p_2) \rightarrow e^-(p_3) e^-(p_4)\ \{\gamma(p_5) \gamma(p_6)\}
\end{equation}
including effects due to the non-vanishing electron mass. 
These corrections are composed of three parts, as illustrated in Figure~\ref{fig:diags}. First, double-real corrections are obtained by integrating the tree-level matrix element (squared) with two additional photons in the final sate over the photon phase space. Second, real-virtual corrections require the interference of the one-loop amplitude and tree-level amplitude with one additional photon. Finally, the double-virtual corrections involve the interference of the two-loop amplitude for $e\,e\to e\,e$ with its tree-level amplitude as well as the corresponding one-loop amplitude squared. These three  individual parts are physically indistinguishable for soft photons and infrared divergent. Only the combination of all contributions leads to a finite, physical result. 

The corrections can be split into purely photonic contributions and fermionic ones that are due to vacuum polarisation (VP). The fermionic part takes into account all three lepton flavours as well as non-perturbative hadronic effects, indicated by the grey blob in Figure~\ref{fig:fermion}. The corresponding method will be presented in the second part of this section after discussing our approach towards photonic corrections first. The double-real contribution of the fermionic part, shown top left in Figure~\ref{fig:fermion} corresponds to the process
\begin{equation}
    e^-(p_1) e^-(p_2) \rightarrow e^-(p_3) e^-(p_4)\ \ell^+(p_5) \ell^-(p_6)
\end{equation}
with additional leptons (or hadrons) in the final state. Since this is a final state with a measurable difference it can be disentangled from \moller{} scattering. It is separately finite if fermion masses are not neglected. Hence, we have chosen not to include the double-real fermionic part in our results presented below. 

The matrix elements are regularised in $d=4-2\epsilon$ dimensions and renormalised in the on-shell scheme. They are implemented in the parton-level integrator \mcmule{} that is based on FKS$^\ell$~\cite{Engel:2019nfw}, a QED extension of the FKS subtraction scheme~\cite{Frixione:1995ms, Frederix:2009Yq} beyond NLO. \mcmule{} allows to consistently perform the phase-space integration in the presence of soft singularities and to calculate physical observables in a fully-differential way. 

\subsection{\label{photonic}Photonic corrections}

All tree-level and one-loop photonic matrix elements were calculated analytically with the full dependence on the electron mass. The diagrams were generated with QGraf~\cite{NOGUEIRA1993279} and calculated with the Mathematica code Package-X~\cite{Patel:2015tea}. While for most of the obtained expressions a sufficiently stable and efficient implementation was possible, a different approach had to be taken in the case of the numerically delicate real-virtual matrix element. In this case next-to-soft stabilisation, previously developed in the context of Bhabha scattering~\cite{Banerjee:2021mty}, was used. This method is based on the observation that the numerical instabilities are strongly enhanced when the emitted photon becomes soft. We have therefore expanded the real-virtual matrix element for small photon energies including the non-universal next-to-soft contribution. This allows to rely on OpenLoops~\cite{Buccioni:2017yxi,Buccioni:2019sur} in its standard mode for the bulk of the phase space and otherwise to switch to the next-to-soft approximation. This approach therefore yields a stable and fast implementation of the problematic real-virtual contribution.

The two-loop matrix element for M\o ller scattering can be derived from corresponding Bhabha results via the crossing relation $p_2 \leftrightarrow -p_4$. However, even though Bhabha scattering is one of the best studied processes in particle physics, the exact two-loop contribution is currently not known. The approximation where the electron mass $m$ is set to zero was calculated some time ago~\cite{Bern:2000ie}. Building upon this result it was later possible to determine the leading mass effects based on the universal factorising structure of collinear divergences~\cite{Penin:2005eh, Mitov:2006xs, Becher:2007cu}. In this approximation power suppressed terms of $\mathcal{O}{(m^2/q^2)}$ are neglected with $q^2$ corresponding to any of the Mandelstam
invariants $s=(p_1+p_2)^2$, $t=(p_1-p_3)^2$, and $u=(p_1-p_4)^2$. Even for low-energy observables this approximation is well justified due to the small electron mass. This massification therefore allows to straightforwardly calculate both the logarithmically enhanced and the constant mass terms based on the massless result. Recently, this procedure was extended to processes that include a heavy mass~\cite{Engel:2018fsb}. In this more general approach a factorisation anomaly~\cite{Becher:2010tm,Becher:2011pf} was discovered in fermionic contributions resulting in the breaking of naive factorisation.
This in turn explains the occurrence of additional powers of large logarithms that result in a larger massification error. We have therefore only used the massification procedure for the photonic two-loop contribution and followed a different approach in the fermionic case.

The analytic continuation that has to be performed after crossing is non-trivial. Fortunately, this step was de facto already performed in the case of the massless two-loop matrix element~\cite{Bern:2000ie} as it was calculated for all three possible kinematic channels for $2 \rightarrow 2$ processes. It was therefore straightforward to construct the massless two-loop matrix element for \moller{} scattering based on this. We then massified the corresponding expression ourselves arriving at the desired result.

We have verified the non-radiative part of our calculation at NLO by comparing to corresponding results in~\cite{Kaiser:2010zz} and found exact agreement. We have further compared to full NLO results provided by~\cite{epstein} using the Monte Carlo code from~\cite{Epstein:2016lpm}. On account of this code being an event generator and not a full dedicated integrator, we found agreement, up to Monte Carlo errors, at the level of precision achieved. To still allow for a robust check of our calculation we have implemented all photonic matrix elements in \mcmule{} such that they can directly be crossed to Bhabha scattering. This then allows us to verify our calculation indirectly by comparing to the much richer literature of Bhabha scattering. In particular, we have compared to the state-of-the-art Monte Carlo generator {\tt BABAYAGA} that is based on a parton shower algorithm matched to the exact NLO result~\cite{Balossini:2006wc}. As presented in~\cite{Banerjee:2021mty}, we have found exact agreement at NLO and a deviation at the level of $17\%$ for the NNLO correction, consistent with the expected precision of the logarithmic approximation from the parton shower.

\subsection{\label{fermionic}Fermionic corrections}

As mentioned above it is not sensible to apply the methods used for photonic corrections to the fermionic contributions because of the occurrence of the factorisation anomaly in the massification procedure. Moreover, at low energies non-perturbative hadronic contributions become relevant. Fortunately, all leptonic and hadronic contributions to \moller{} scattering are due to VP insertions. We can therefore calculate all of them simultaneously by including all contributions in the VP
\begin{equation}
\Pi = \Pi_e+\Pi_\mu+\Pi_\tau+\Pi_\mathrm{had}.
\end{equation}
The one-loop contribution to the leptonic VP can be calculated trivially and the two-loop result can be extracted from~\cite{Djouadi:1993ss}. For the non-perturbative hadronic VP, on the other hand, we rely on the Fortran library {\tt alphaQED}~\cite{Jegerlehner:2001ca,Jegerlehner:2006ju,Jegerlehner:2011mw} that calculates $\Pi_\mathrm{had}$ based on experimental data.

The full photon propagator
\begin{equation}\label{eq:photonprop}
    \frac{-i g^{\mu\nu}}{q^2\big(1-\Pi(q^2)\big)}
    =\frac{-i g^{\mu\nu}}{q^2}
    \Big(1+\Pi(q^2)+\Pi^2(q^2) + ... \Big)
\end{equation}
automatically resums VP insertions to all orders in perturbation theory. In this paper, however, we take a strict fixed-order approach for all contributions. In what follows we therefore use the expanded version of the photon propagator given on the r.h.s. of~\eqref{eq:photonprop}. The impact of missing higher-order effects -- together with resummation of soft and collinear photon emission -- is left for future studies.

For a large class of diagrams, such as the two diagrams on the right of Figure~\ref{fig:fermion}, the VP contributions factorise and can thus be calculated straightforwardly based on photonic NLO matrix elements. The verification of the photonic corrections described in Section~\ref{photonic} therefore also presents a strong check in this case. The calculation of the non-factorisable vertex and box diagrams, however, is more involved. Traditionally these contributions are calculated dispersively, see e.g.~\cite{Fael:2019nsf}. In our case we have instead used the hyperspherical formalism~\cite{Levine:1974xh,Levine:1975jz} where it was possible to reuse many of the results of the analogous calculation in muon-electron scattering~\cite{Fael:2018dmz}. In this approach the non-factorisable two-loop contribution is written as an integral over the radial part of the Wick-rotated loop momentum
\begin{equation}
    \mathcal{M}_\mathrm{VP}^\mathrm{non-fac} =
    \int_0^\infty dQ^2 \Pi(-Q^2)  K(Q^2,s,t,u)
\end{equation}
which can be performed numerically. The kernel function $K$ results from the analytic integration over the solid angle of the loop momentum. In the case of triangles this integration can be elegantly performed after writing the propagators in terms of Gegenbauer polynomials and making use of the corresponding orthogonality relation. For box integrals, however, this method does not work and the corresponding calculation is therefore more complicated. In this case the general result can be found in~\cite{Laporta:1994mb}. 

The main advantage of the hyperspherical approach compared to the dispersive calculation is that the VP is integrated over space-like momenta avoiding hadronic resonances. Furthermore, the threshold singularities are transformed into logarithmic divergences in the kernel functions through the angular integration. Nevertheless, the numerical integration over $Q^2$ has to be performed with care. Since large cancellations can occur in the kernel functions, they have to be expanded in these problematic regions. In particular, this is the case for $Q^2\rightarrow 0$, $Q^2\rightarrow \infty$, and around threshold singularities.

We have verified our calculation of the non-factorisable fermionic contributions by comparing to corresponding crossed results for Bhabha scattering. First of all, we have obtained complete agreement in the case of closed electron loops with the exact analytic calculation in~\cite{Bonciani:2004gi}. Furthermore, we have compared the muon and the tau contributions to approximate results from~\cite{Actis:2007gi} that neglect terms of $\mathcal{O}(m_f^2/q^2)$ with $m_f$ the mass of any of the fermions and $q^2\in\{s,t,u\}$. As expected we do not find exact agreement but instead observe the correct converging behaviour when the energy scale $q^2$ is increased.

\section{\label{sec:results}Results}

With the contributions discussed in the previous section implemented in \mcmule{} we can compute any infrared safe observable of \moller{} scattering at NNLO in QED.  

To illustrate this, as a first application we use a kinematic set up tailored to the PRad~II experiment. We consider an electron beam with energy $E_b = 1.4\ \mathrm{GeV}$ incident on a target electron at rest. We refer to the outgoing electron with the smaller (larger) scattering angle as the `narrow' (`wide') electron. The corresponding scattering angles are denoted by $\theta_n$ and $\theta_w$, respectively. We further define the inelasticity $\eta = E_b +m - E_n - E_w$ and the coplanarity $\zeta = \big{|}180^{\circ}-|\phi_n-\phi_w|\big{|}$ with $E_i$ and $\phi_i$ the energy and azimuth of the narrow and wide electrons. Both quantities are zero for elastic events and can therefore be used to restrict hard photon emission. We approximate the experimental design with the cuts
\begin{equation}\label{eq:fiducial}
    0.5^{\circ} < \theta_n, \theta_w < 6.5^{\circ}, \quad
    \eta < 3.5\sigma_E, \quad
    \zeta < 3.5\sigma_\phi,
\end{equation}
with $\sigma_E=37.7\ \mathrm{MeV}$ and $\sigma_\phi = 2.1^{\circ}$ the expected detector resolutions for the considered beam energy~\cite{prad}. In addition to this fiducial observable we also consider a simpler version that does not restrict hard photon radiation. To this end we convert the angular cuts in~\eqref{eq:fiducial} to corresponding restrictions for the $t$- and $u$-channel momentum transfers using the kinematic relations for elastic scattering. The corresponding cuts for this simplified scenario then read
\begin{equation}\label{eq:simple}
    -1295\ \mathrm{MeV}^2 \lesssim t,u \lesssim -135\ \mathrm{MeV}^2.
\end{equation}

\begin{figure}[t]
\centering
 \begin{tabular}{c|r r|| r r} 
 & \multicolumn{2}{c||}{$\sigma/\rm \upmu b$} & \multicolumn{2}{c}{$\delta K^{(i)}/\%$} \\
  & \multicolumn{1}{c}{\tt fiducial} & \multicolumn{1}{c||}{\tt simple} & \multicolumn{1}{c}{\tt fiducial} & \multicolumn{1}{c}{\tt simple} \\[0.3ex] 
 \hline
 \rule{0pt}{3ex}
 $\sigma^{(0)}$ & \tt 2291.02  &\tt 2291.02& & \\[0.3ex]
 \hline
 \rule{0pt}{3ex}
 $\sigma^{(1)}_\gamma$ & \tt   -148.36  &\tt  78.23& \tt -6.476&\tt 3.415\\
 $\sigma^{(1)}_\mathrm{VP}$ & \tt   19.33  &\tt  19.33& \tt 0.844&\tt 0.844\\[0.5ex]
 \hline
 \rule{0pt}{3ex}
 $\sigma^{(2)}_\gamma$ & \tt  2.82  &\tt  -0.17& \tt0.131&\tt  -0.007 \\
 $\sigma^{(2)}_\mathrm{VP}$ & \tt  -1.31  &\tt  -0.06& \tt-0.061&\tt  -0.002 \\[0.5ex]
 \hline\hline
 \rule{0pt}{2.5ex}
 $\sigma_{2}$   & \tt 2163.50  &\tt 2388.35& & \\
\end{tabular}
\captionof{table}{\label{tab:xsection}
The integrated cross section for the fiducial and the simplified cuts at LO,
NLO, and NNLO. All digits are significant compared to the numerical integration error.}
\end{figure}

The order-by-order contributions, $\sigma^{(i)}$, to the integrated
cross section, $\sigma_2=\sigma^{(0)} + \sigma^{(1)} + \sigma^{(2)}$, for both the fiducial and the simplified cuts are presented in Table~\ref{tab:xsection}. The photonic and fermionic corrections are given separately and are denoted by $\sigma^{(i)}_\gamma$ and $\sigma^{(i)}_\mathrm{VP}$, respectively. Additionally, we show the
corresponding $K$ factors defined as
\begin{equation}\label{eq:kfac}
    K^{(i)}=1+\delta K^{(i)} =
    \frac{\sigma_{i}}{\sigma_{i-1}}.
\end{equation}
As expected, the results for the fiducial and the simple cuts agree for the elastic corrections $\sigma^{(0)}$ and $\sigma^{(1)}_\mathrm{VP}$.
We further observe large NLO contributions compared to the naive counting of the expansion parameter $\alpha/\pi$.
The NNLO $K$ factors, on the other hand, are much smaller and in agreement with the naive prediction. For the total cross section we can thus conclude that the NNLO result yields a robust prediction for the considered observable with missing higher-order corrections well under control. Comparing the fiducial and simple results reveals that the inelasticity and coplanarity cut have a noticeable impact on the higher-order corrections. If hard photon emission was even more severely restricted, the NNLO corrections, too, would become large and the resummation of corresponding large logarithms would be required.

\begin{figure}[t]
\centering
\subfloat[Energy distribution of the `narrow' electron]{
         \includegraphics[width=.465\textwidth]{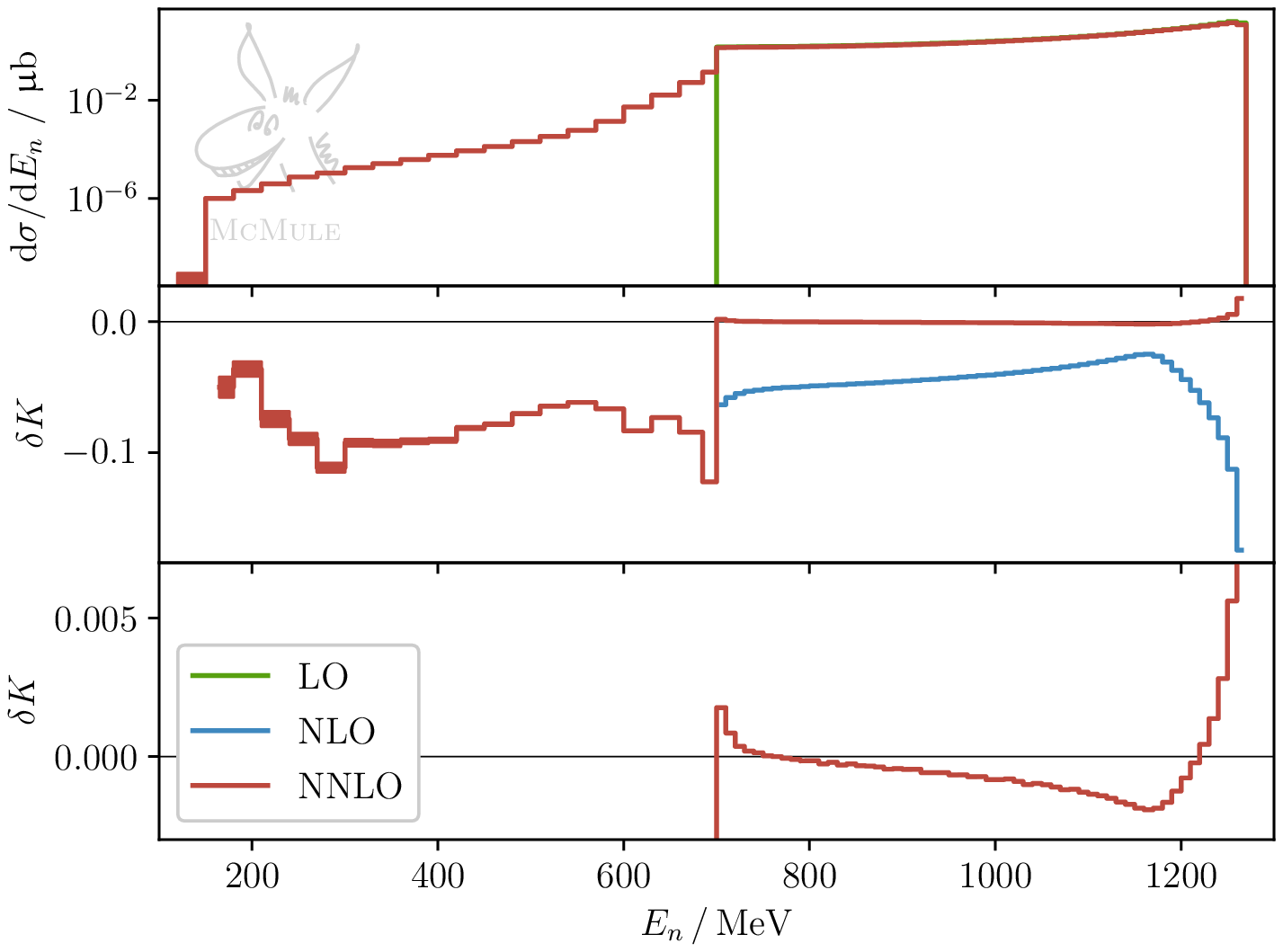}
         \label{fig:energy}
}\\
\subfloat[Angular distribution of the `narrow' and the `wide' electron]{
         \includegraphics[width=.465\textwidth]{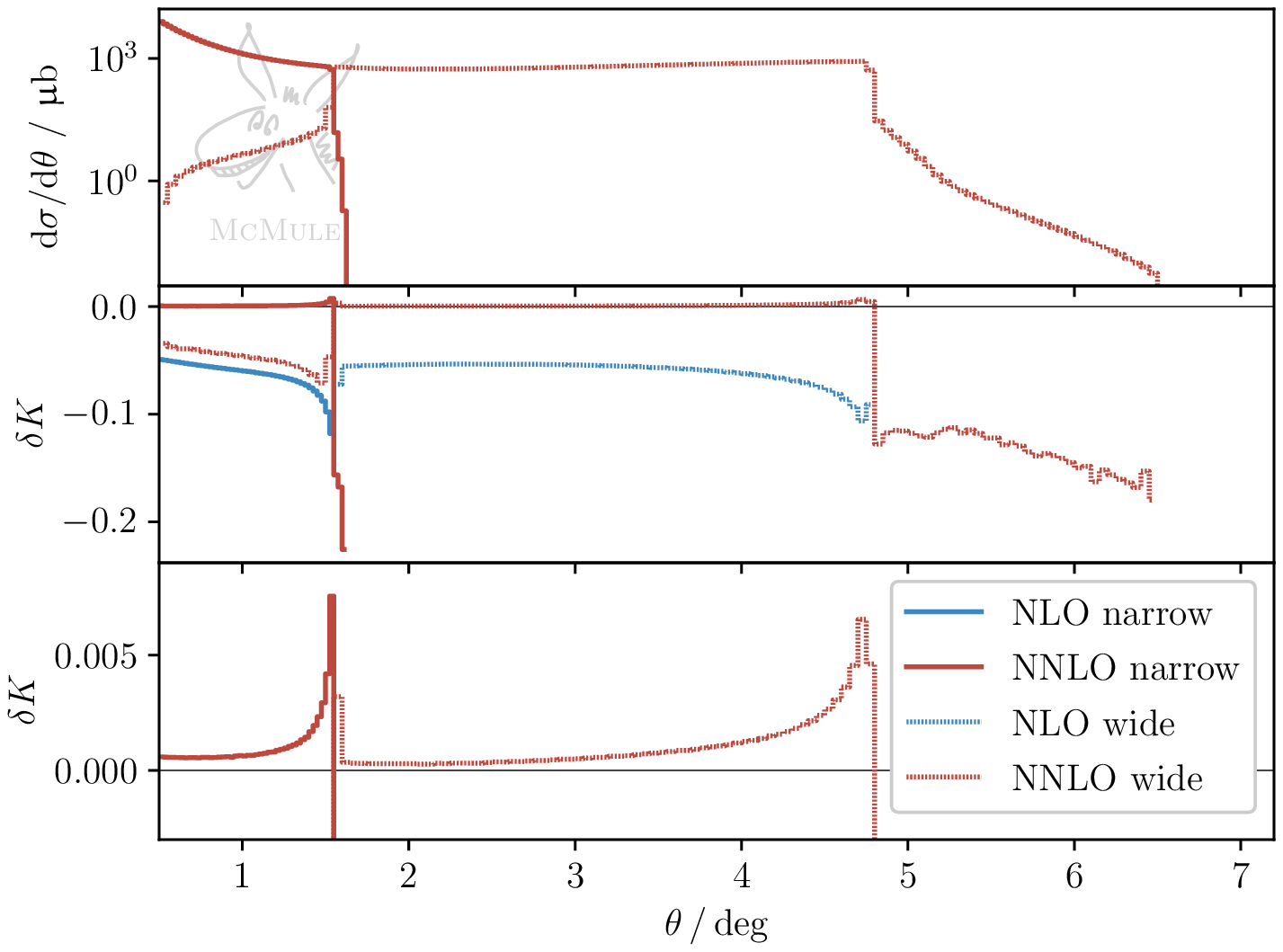}
         \label{fig:theta}
}
\captionof{figure}{\label{fig:distributions}
Differential cross section for the fiducial observable defined in~\eqref{eq:fiducial}.}
\end{figure}

For the more realistic fiducial observable we present differential distributions in Figure~\ref{fig:distributions}. Figure~\ref{fig:energy} shows the corresponding results with respect
to the energy of the `narrow' electron. The angular distributions for both the `narrow' and `wide' electron are given in Figure~\ref{fig:theta}. For both figures the differential cross
section is displayed in the upper panel. In
addition, the middle panel shows the differential version of the $K$ factor defined in~\eqref{eq:kfac}. In regions where the LO cross section is zero the NNLO contribution effectively corresponds to an NLO correction resulting in a large $K$ factor. For this reason the lower panel zooms into the other region where the NNLO $K$ factor is small. In this case we observe a similar behaviour as for the total cross section with large NLO and small NNLO corrections. Contrary to the integrated case, however, the NNLO calculation is indispensable for distributions in order to obtain a reliable prediction also in the region where the LO contribution is zero.

We have presented integrated and differential cross sections for only one possible scenario for PRad~II. Other configurations of interest to PRad and PRad~II will be studied in the future and made available on
\begin{lstlisting}[language=bash]
  (*@\url{https://mule-tools.gitlab.io/user-library}@*)
\end{lstlisting}
The data that enter the results presented in this section can also be found there.

\section{\label{sec:conclusion}Conclusion}

We have calculated the full set of NNLO QED corrections to \moller{} scattering including photonic, leptonic, and non-perturbative hadronic corrections. The fermionic part was calculated with a semi-numerical approach using the hyperspherical formalism without any approximation. The photonic two-loop matrix element, on the other hand, was computed based on the crossed massless Bhabha result via massification resulting in a parametrically small error of $\mathcal{O}(\alpha^2 m^2/q^2)$ relative to LO. Apart from the calculation of the two-loop matrix element, the main challenge in the computation of the photonic corrections is the numerical stability of the real-virtual matrix element. We have obtained a fast and stable implementation of this delicate contribution using a combination of OpenLoops and the next-to-soft stabilisation method recently developed in the context of Bhabha scattering.

All matrix elements were implemented in the parton-level Monte Carlo integrator \mcmule{}. While not an event generator, this framework allows to compute arbitrary fully-differential observables by consistently removing soft divergences originating in the phase-space integration. As a first application of our NNLO calculation we have generated integrated and differential cross sections relevant for the planned PRad~II experiment.
For the total cross section we obtain large NLO and small NNLO corrections ensuring a high-precision prediction. In the case of differential distributions the situation is more subtle. In the absence of a LO contribution the NNLO correction becomes essential to achieve a moderate precision of about $10\%$. To improve this theory accuracy further missing higher-order corrections could be approximated with a parton shower.

\subsection*{Acknowledgement}

We are grateful to A.~Antognini for calling to our attention the necessity
of the calculation and assuring us that anyone can do it. We thank the PRad collaboration for providing the kinematical configuration relevant for the PRad~II experiment. We are indebted to C.~Carloni Calame for engaging in a
detailed comparison with {\tt BABAYAGA}. We also thank C.~Epstein for the comparison with his NLO event generator. It is further a pleasure to thank M.~Fael for his advice regarding the hyperspherical formalism and for helping us verify the corresponding results. We are also grateful to M.~Zoller for his help in using OpenLoops. Our thanks are further extended to P.~Mastrolia and R.~Bonciani for providing the analytic results for the massive fermionic two-loop contributions in electronic form. Last but not least, we thank L.~Dixon for a digital version of the massless two-loop matrix element. PB and TE acknowledge support by the Swiss National Science Foundation
(SNF) under contract 200021\_178967. PB further acknowledges support by the European Union's Horizon 2020 research and
innovation program under the Marie Sk\l{}odowska-Curie grant agreement No
701647. YU acknowledges support by the UK
Science and Technology Facilities Council (STFC) under grant ST/T001011/1.

\bibliographystyle{JHEP}
\small\bibliography{moller}

\end{document}